\documentclass[aps,showpacs,onecolumn,superscriptaddress,12pt]{revtex4-1}
\usepackage{graphicx,tikz,amsmath,amssymb}
\usepackage{psfrag,subfigure}

\renewcommand{\Re}{\textrm{Re}}

\newcommand{\eps}{\varepsilon}

\begin{document}

\title{Scattering of Evanescent Wave by Periodic Array of Nanowires}

\author{Leonid L. Frumin}
\affiliation{Institute of Automation and Electrometry, Siberian Branch, Russian Academy of Sciences, Novosibirsk 630090, Russia}

\affiliation{Novosibirsk State University, Novosibirsk 630090, Russia}

\author{Anton V. Nemykin}
\affiliation{Institute of Automation and Electrometry, Siberian Branch, Russian Academy of Sciences, Novosibirsk 630090, Russia}
\affiliation{Novosibirsk State University, Novosibirsk 630090, Russia}

\author{Sergey V. Perminov}
\affiliation{A.V. Rzhanov Institute of Semiconductor Physics, Siberian Branch, Russian Academy of Sciences, Novosibirsk 630090, Russia}

\author{David A. Shapiro}
\affiliation{Institute of Automation and Electrometry, Siberian Branch, Russian Academy of Sciences, Novosibirsk 630090, Russia}

\affiliation{Novosibirsk State University, Novosibirsk 630090, Russia}

\address{Corresponding author: shapiro@iae.nsk.su}

\begin{abstract}
The scattering of electromagnetic wave by a periodic array of nanowires is calculated by the boundary element method. The method is extended to the infinite grating near the interface between two dielectrics. A special Green function is derived that allows to study the evanescent wave. The Rayleigh--- Wood's anomalies
are found in the period-to-wavelength dependence of the average Pointing vector in the wave zone. For thin wires the calculations are shown to agree with the two-dimensional coupled dipole approximation.

\end{abstract}

\maketitle

\section{Introduction}

The travelling wave falling onto the boundary between two
dielectrics by the angle of total internal reflection generates an
evanescent wave in the second medium. The illumination by
evanescent wave is widely used in applications. In near-field
scanning optical microscopy it allows one to detect the desired
signal with no background \cite{aplPDL84}. Total internal
reflection fluorescence, exploiting decay of evanescent wave,
achieves nanometer resolution along the optical path through a
thin object \cite{prlWRVS10}. Covering the fiber surface by
nanowires improves the sensitivity of fiber Bragg grating
refractometry \cite{ntB12}.

While the scattering problem for evanescent wave is of importance,
it has, however, no analytical solution. The necessity of treating
the evanescent wave along with its source exaggerates the
calculation. For travelling wave the analytical solution exists
for the homogeneous medium and one circular cylinder
\cite{Harrington61}. The numerical methods allows one to solve the
problem for one elliptic cylinder \cite{jtpGZ06}. The numerical
solution for two different dielectric half spaces are found
recently for one \cite{olBFPS11} and two \cite{eplBPFS12}
cylinders. The scattering by few cylinders has very small
cross-section and then is difficult to observe. To enhance the
effect one needs to consider a number of cylinders, but growing
the number of scatterers results in additional computational
difficulty. To remedy this contradiction between the observation
needs and computation difficulties, we reduce a two-dimensional
infinite periodic array to only one unit cell with the help of
periodicity. The aim of the present paper is to calculate the
scattering of evanescent wave by the periodic array of parallel
cylinders of arbitrary diameter.

The periodic structures of scatterers with subwavelength dimension
of each enhance the signal compared to one scatterer
\cite{prpBFL07,RevModPhys.79.1267,jnFEB11}. Two-dimensional
structures are important, for instance, for photonic crystals
\cite{PhysRevLett.108.037401} or metamaterials \cite{olSDS05}. The
diffraction grating can be characterized by its Rayleigh ---
Wood's anomalies. We consider these resonances and compare their
structure with the case of travelling wave and with widely used
coupled dipole approximation.

For this purpose we propose a modification of the boundary
integral equations, Section \ref{Method}, which are the basis of
the Boundary Element Method (BEM) \cite{Brebbia84,Wu-00,Beer08}.
We derive a special Green function that takes into account the
evanescent wave together with its source, the dielectric
half-space, in Subsection \ref{Green}. The method is extended to
the periodic array in Subsection \ref{BEM}. In Section
\ref{Results} we present the results of calculation. Subsection
\ref{Raylegh} contains the study of average energy flux in the
far-field domain as a function of period-to-wavelength ratio,
including splitting of Rayleigh--- Wood's anomalies. The
Subsection \ref{Dipole} is devoted to the calculation within
dipole approximation, in comparison to the more rigorous BEM, for
the travelling exciting wave. The coupled dipoles equations are
presented for both finite and infinite periodic grating.

\section{Method}\label{Method}
\subsection{Green Function}\label{Green}

Let us consider two media, 1 and 2, separated by a plane boundary at
$y=0$. We look for the specific Green function $G(x,y;x',y')$
satisfying inhomogeneous equation
\begin{equation}\label{Poisson}
\bigtriangleup G+k^2G=\delta(x-x')\delta(y-y')
\end{equation}
in these media. Function $G$ depends on the difference $x-x'$ only
due to translational symmetry. The boundary condition at $y=0$ for $p$-wave is
\begin{equation}\label{BoundaryG}
\left[G(x,y;x',y')\right]_{y=0}= \left[\frac1\eps\frac{\partial
G(x,y;x',y')}{\partial y}\right]_{y=0}=0.
\end{equation}

After the Fourier transformation
\begin{equation}\label{Fourier}
G(x,y;0,y')=\frac1{2\pi}\int_{-\infty}^\infty
G_q(y,y')e^{iqx}\,{dq}
\end{equation}
Eq. (\ref{Poisson}) is reduced to an ordinary differential
equation having exponential solutions. Using conditions
(\ref{BoundaryG}) we obtain the function. The expression for $y'>0$ in $q$-domain is
\begin{equation}\label{Green-Fourier}
G_q=-\frac1{2\mu_2}\begin{cases}
\left[1+r(q)\right]e^{\mu_1 y-\mu_2 y'},&y<0,\\
e^{-\mu_2|y-y'|}+r(q)e^{-\mu_2(y+y')},
& y>0.
\end{cases}
\end{equation}
Here
\begin{equation}\label{TM}
r(q)=\frac{\eps_1\mu_2-\eps_2\mu_1}{\eps_1\mu_2+\eps_2\mu_1}
\end{equation}
is the Fresnel reflection coefficient of $p$-wave at normal incidence
\cite{BW65}, $\mu_{1,2}^2=q^2-k_{1,2}^2$. Carrying out
Fourier transformation (\ref{Fourier}) of function
(\ref{Green-Fourier}) at $y>0$ we have two terms
\begin{eqnarray}\label{first-Green}
G_1=-\int\limits_{-\infty+i0}^{\infty-i0}
e^{-\mu_2|y-y'|+iq(x-x')}\frac{dq}{4\pi\mu_2},\\
G_2=-\int\limits_{-\infty+i0}^{\infty-i0}
e^{-\mu_2(y+y')+iq(x-x')}\frac{r(q)\,dq}{4\pi\mu_2},
\label{second-Green}
\end{eqnarray}
where $G=G_1+G_2$. The sign of the square root is given by the rule
$\sqrt{q^2-k_2^2}\to-i\sqrt{k_2^2-q^2}, q^2<k_2^2$.

The first term (\ref{first-Green}) can be calculated analytically and reduces
to the known Green function of Helmholtz equation in the homogeneous space
\begin{equation}\label{Fundamental}
G_1(\vec{r};\vec{r}\,')=\frac1{4i}{\mathcal{H}_0^{(1)}
\left(k_2\rho_-\right)},
\end{equation}
where $\mathcal{H}_0^{(1)}$ denotes the Hankel function of the first
kind \cite{Olver10},
\begin{equation}\label{rho}
\rho_{\pm}=\sqrt{(x-x')^2+(y\pm y')^2}
\end{equation}
are the distances from the source and its image.
The second term $G_2$ gives the effect of the reflected image source. The
amplitude of the source at each $q$ is equal to the reflection
coefficient $r(q)$. Thus along with the point source at $(x',y')$ we
have to consider mirror-image source $r(q)$ at $(x',-y')$. The
total field at the upper half-plane is the sum of the fields
generated by the two sources at each $q$. This approach
intrinsically takes into account the multiple scattering. The Green
function of this type was studied for homogeneous waves: spherical
acoustic, see \cite{B80,LL6}, or cylindrical electromagnetic waves
\cite{Pincemin94}. A general approach has been built, based on
integral equation, to study the field scattered by an object near
two-media interface, for excitation by homogeneous \cite{ocG89} and
evanescent \cite{olBFPS11,eplBPFS12} light wave.

Analogous calculations give the Green function for the source inside
the lower dielectric.
\begin{equation}\label{hidden}
G_q=-\frac1{2\mu_1}\begin{cases}
e^{-\mu_1|y-y'|}-r(q)e^{\mu_1(y+y')},
& y<0,\\
\left[1-r(q)\right]e^{-\mu_2 y+\mu_1 y'},& y>0.
\end{cases}
\end{equation}
Comparing function (\ref{hidden}) with (\ref{Green-Fourier}) we see that the expression can be obtained by the symmetry transformation, namely, changing medium indices $1\leftrightarrow2$ and the sign of the ordinate
$y\to-y,y'\to-y'$.
The case of $s$-wave the boundary conditions (\ref{BoundaryG}) are
replaced by
\begin{equation}\label{BoundaryTE}
\left[G(x,y;x',y')\right]_{y=0}= \left[\frac{\partial
G(x,y;x',y')}{\partial y}\right]_{y=0}=0.
\end{equation}
The result can be also obtained from (\ref{Green-Fourier}) replacing the
Fresnel reflection coefficient (\ref{TM}) for normal incidence by its expression for $s$-wave
\begin{equation}\label{TE}
r(q)=\frac{\mu_2-\mu_1}{\mu_1+\mu_2}.
\end{equation}

The compound Green function is valid in both half-spaces and makes it possible to solve the scattering problem near the interface between two dielectrics. This function  is yet for few cylinders but worthless for the periodic grid. In the next section we explain how to extend the computation for the infinite grid with the help of Floquet theorem and derive the corresponding effective Green function.

\subsection{Boundary Element Method}\label{BEM}

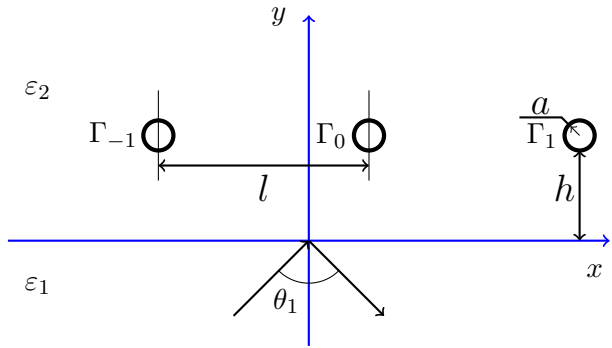
\begin{figure}\centering
\begin{tikzpicture}[scale=4]
\draw[thick,blue,->](-1,0)--(1,0);
\draw[thick,blue,->](0,-0.35)--(0,0.75);
\draw[thick,<->] (-0.5,0.25) -- (0.2,0.25);
\draw (0.95,-.1) node {$x$};
\draw (-.1,0.745) node {$y$};
\draw (-0.9,-.15) node {$\eps_1$};
\draw (-0.9,.5) node {$\eps_2$};
\draw (-0.075,-0.2) node {$\theta_1$};
\draw (-0.15,0.175) node {\large $l$};
\draw[thick,<->] (0.9,0)--(0.9,0.3);
\draw (0.85,0.175) node {\large $h$};
\draw (0.2,0.2)--(0.2,0.5);
\draw (-0.5,0.2)--(-0.5,0.5);
\draw[ultra thick](0.9,0.35) circle (0.05);
\draw[ultra thick](0.2,0.35) circle (0.05);
\draw[ultra thick](-0.5,0.35) circle (0.05);
\draw (0.775,0.35) node {$\Gamma_1$};
\draw (0.075,0.35) node {$\Gamma_0$};
\draw (-0.65,0.35) node {$\Gamma_{-1}$};
\draw[->] (0.9,0.35) -- (0.87,0.38);
\draw[thick] (0.87,0.38) -- (0.84,0.41) -- (0.7,0.41);
\draw (0.77,0.45) node {\large $a$};
\draw[thick,->] (-0.25,-0.25)--(0,0);
\draw[thick,->] (0,0)--(0.25,-0.25);
\draw (-0.1,-0.1) arc (225:315:0.141);
\end{tikzpicture}
\caption{A cross-section of the periodic array of cylinders: $a$ is the radius, $h$ is the gap, $l$ is
the period, $\theta_1$ is the light incidence angle.}\label{fig:scheme}
\end{figure}

Let us consider a scattering of a plane wave by the periodic grid
of identical parallel cylinders with dielectric permittivity
$\eps_3$, see Fig. \ref{fig:scheme}. The grid is located in
dielectric half-space 2 (with dielectric permittivity $\eps_2$)
parallel to its boundary with half-space 1 ($\eps_1$). A
travelling wave with frequency $\omega$ falls from half-space 1 by
the angle of incidence $\theta_1$. Denote
$k_{1,2}=k_0\sqrt{\eps_{1,2}}$ the wavenumber in each dielectric,
where $k_0=\omega/c$, $c$ is the speed of light. As known from
Fresnel formulas, the transmitted wave in media 2 becomes
evanescent when angle $\theta_1$ exceeds the angle of total
internal reflection. Choose axis $z$ along the axes of cylinders,
axis $x$ along the boundary, and axis $y$ perpendicular to the
boundary, as shown in Fig. \ref{fig:scheme}.

We concentrate on the p-wave (TM case), when the magnetic field has only
$z$-component, which consists of the falling and the scattered
fields: $H_z=H_0+H'$. With the help of integral Green formula we
write the coupled boundary integral equations for scattered field
$u=H'$ and its normal derivative $f= {\partial H'}/{\partial n}$ at a
contour point $P=(x,y)\in\Gamma$:
\begin{equation}\label{BEM1}
-\frac{u(P)}{2}+\int_{\Gamma}\left[\frac{\partial
G_{\textrm{out}}}{\partial n_{Q}} u(Q)-G_{\textrm{out}}f(Q)\right] d
Q=0.
\end{equation}
Here $n_Q$ is the internal normal in point $Q=(x',y')\in\Gamma$,
$G_{\textrm{out}}(P,Q)$ is the external Green function. The
integration point $Q$ runs through compound contour
$\Gamma=\cup_m\Gamma_m$, where $\Gamma_m$ is the contour of $m$-th
circle (cross section of cylinder): $m=1,2,\dots,N$; $N$ is the
number of cylinders. The integral over an unlinked domain is a sum of
integrals over partial contours $\Gamma_m$. This equation corresponds
to the outer limit, i.e. the limit of Green formula when the
observation point tends to contour $\Gamma$ from the external side.
For the outer limit we use the special Green function described in
Sec. \ref{Green}. It explicitly takes into account the boundary
conditions at the plane interface between the dielectrics 1 and 2, and
has a form of à divergent wave at infinity \cite{olBFPS11,eplBPFS12}.

The second integral equation arises as a result of the passage to the inner
limit, i.e. when point $P$ tends to the contour from internal area:
\begin{eqnarray}\label{BEM2}
\frac{u(P)}{2}+\int_{\Gamma_m}\left[\frac{\partial
G_{\textrm{in}}}{\partial n_{Q}} u(Q)-G_{\textrm{in}}f(Q)\right] d Q
=B_0, \\
B_0 = -\frac{u_0}{2} - \int_{\Gamma_m}\left(\frac{\partial
G_{in}}{\partial n}u_0 - G_{in} f_0\right) d Q.
\end{eqnarray}
Here $G_{\textrm{in}}(P,Q)$ is the internal Green function, $u_0=H_0, f_0=\partial H_0/\partial n$. For internal domain we use the cylindrical wave
\begin{eqnarray}
G_{\textrm{in}}(x,y;x',y')=\frac1{4i}\mathcal{H}^{(1)}_0(k_3\rho_-).
\end{eqnarray}

For $N$ cylinders there are $N$ internal and $N$ external equations.
Solving the set of $2N$ equations we find functions $u(P),f(P),
P\in\Gamma$. If we know functions $u,f$ at the contour, then
scattered field in arbitrary external point $R=(x,y)\notin\Gamma$ can
be found from the Green theorem
\begin{equation}\label{GreenFormula1}
u(R)=\int_{\Gamma}\left[\frac{\partial
G_{\textrm{out}}}{\partial n_{Q}}
u(Q)-G_{\textrm{out}}f(Q)\right] \,dQ,
\end{equation}
where $G_{\textrm{out}}\equiv G_{\textrm{out}}(R,Q)$ is the external Green function.

Eqs. (\ref{BEM1}), (\ref{BEM2}) refer to the case of a finite
number $N$ of the cylinders. Passing to the limit $N\to\infty$ we
consider points at the compound contour with equal $y_m=y_0$ and
$x_m=x_0+ml; m=0,\pm1,\pm2,\dots$  A zeroth cylinder $m=0$ is chosen arbitrary
for the infinite grating. The following property (also known as
Floquet theorem) is a consequence of the periodicity:
\begin{equation}\label{BlochFloke}
u(x_m,y_m)=u(x_0+m d,y_0)=u(x_0,y_0)e^{i k_\bot m l},
\end{equation}
where $k_\bot=k_1\sin\theta_1$ is the tangential component of the wavevector conserved at the boundary. Taking into account the same property of function $f(Q)$ and reversing the order of summation and integration in Eq. (\ref{BEM1}) and (\ref{GreenFormula1})\footnote{We assume the absolute convergence of the corresponding seria.} we obtain the sum:
\begin{equation}\label{Sum1}
\sum_{m=-\infty}^{+\infty}
\left[\frac{\partial G_{\textrm{out}}}{\partial n_{Q}}
u(Q)-G_{\textrm{out}}f(Q)\right]e^{i k_\bot m l}.
\end{equation}

The special Green function (\ref{Fourier}), (\ref{Green-Fourier}) in
the upper half space 2 can be written as a contour integral
\begin{equation}\label{Integral-representation}
G_{\textrm{out}}=-\int\limits_C
\left(e^{-\mu|y-y'|}+r_ke^{-\mu(y+y')}\right)\frac{e^{ik\Delta}dk}{4\pi\mu},
\end{equation}
where $y,y'>0, \Delta=x-x'$, $G_{\textrm{out}}\equiv G_{\textrm{out}}(x,y;x',y')$,
$\mu=\sqrt{k^2-k^2_{2}}$, $C=(-\infty+i0,+\infty-i0)$. Consider the
latter sum in Eq. (\ref{Sum1})
\begin{equation}\label{sumG}
G_{\textrm{eff}}=\sum_{m=-\infty}^{+\infty}G_{\textrm{out}}(x,y;x'_m,y'_m)e^{i
k_\bot ml}.
\end{equation}
From (\ref{sumG}) and integral representation
(\ref{Integral-representation}), alternating the order of summation
and integration, we obtain the periodic delta function
\begin{equation}\label{DeltaFunc}
\sum_{m=-\infty}^{+\infty}e^{-i k m d +i k_\bot m
d}=\frac{2\pi}{l}\delta\left( k-k_\bot-n \varkappa\right),
\end{equation}
where $n=0,\pm1,\pm2,\dots,$ $\varkappa=2\pi/l$ is the reciprocal
lattice vector. Then integrating the delta-function over $k$ we get
the expression for $G_{\textrm{eff}}$
\begin{eqnarray}\label{I2}
G_{\textrm{eff}}=-\sum_{n=-\infty}^{+\infty}
\frac{e^{i(k_\bot+\varkappa
n)\Delta_0}}{2l\mu_{n}}
\Bigl[e^{-\mu_{n}|y-y'_0|}+r_n e^{-\mu_{n}(y+y'_0)}\Bigr],
\end{eqnarray}
where $\Delta_0=x-x'_0$, $\mu_{n}=\sqrt{(k_\bot+n
\varkappa)^2-k^2_{2}}$, $r_n=r_{k_\bot+n \varkappa}$. The former term in
Eq. (\ref{Sum1}) is reduced to the normal derivative of (\ref{I2})
analogously, changing the order of normal derivative and summation.

Expression (\ref{I2}) determines the effective Green function, taking
into account the infinite grid as a whole. If we replace
$G_{\textrm{out}}$ by $G_{\textrm{eff}}$ in external integral
equation (\ref{BEM1}) and Green theorem (\ref{GreenFormula1}), then
the path of integration will be $\Gamma_0$. It is the zeroth circle,
the connected contour, instead of infinite unlinked contour $\Gamma$.
Only one internal equation should be solved, for $m=0$. After finding
functions $u(P), f(P)$  we use Eq. (\ref{GreenFormula1}) for
calculation of the field in the wave zone.

\section{Results}\label{Results}

\subsection{Rayleigh --- Wood's anomalies}\label{Raylegh}

We study the transparency of the grating as a function of distance
$l$ between the neighbor cylinders. The transparency is defined as
a ratio of $y$-projection of average Pointing vector to the energy
flux density of the falling wave. The Pointing vector is averaged
over the period and calculated at sufficiently large distance from
the boundary $y\sim 10\lambda$, where $\lambda=1.5\ \mu$m is the
fixed wavelength.

\begin{figure}
\psfrag{abs}{$l/\lambda$}
\psfrag{ord}{\hspace{-0.5cm}Transparency}

\includegraphics[width=\columnwidth]{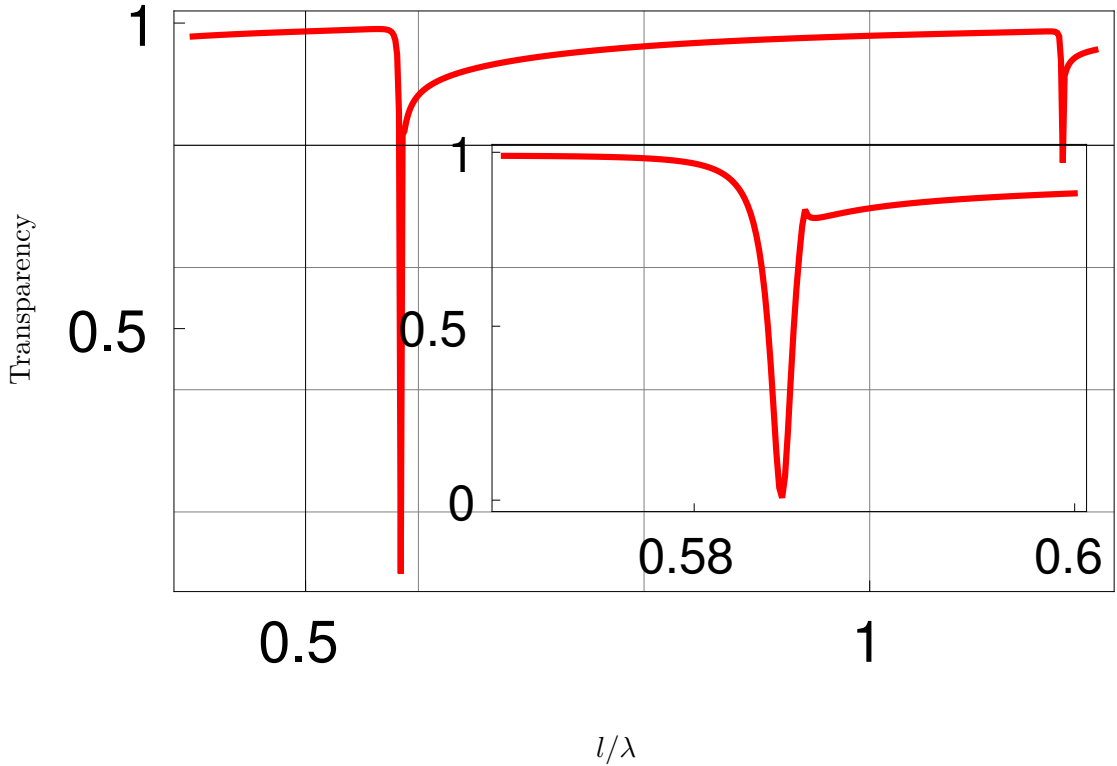}
\caption{Raleigh --- Wood's anomalies in the transparency for
travelling wave ($\eps_1=1$, the cylinders' radii $a=0.1\ \mu$m).
The first dip with higher resolution is shown in the
inset.}\label{f:uniform}
\end{figure}

Only terms corresponding to travelling waves with $(k_\bot+n
\varkappa)^2<k^2_2$ (and then $\Re\mu_n=0$) have to be kept for
calculations in the wave zone. Thus, the sum (\ref{I2}) becomes
finite and relatively easy for numerical calculation compared to
integral representation (\ref{Integral-representation}). Fig.
\ref{f:uniform} shows the transparency as a function of the
spacing between the cylinders, $l$, (rated to the wavelength) for
the travelling wave ($\eps_1=\eps_2=1, \theta_1=45^\circ$). The
grating is placed close to the surface: $h=0.01\ \mu$m.

We see the narrow dips, so-called Rayleigh --- Wood's anomalies
(RWA). These resonances take place when the grazing mode appears.
The condition of resonance is synchronization of the phases
between the waves scattered by neighbor cylinders. The grazing
modes may have both positive and negative order
$n=\pm1,\pm2,\dots$ The positive orders $n>0$ generate the surface
wave going in negative direction and vice versa. For
positive/negative mode the synchronism condition is
\begin{equation}\label{synchronism}
    k_1l\sin\theta_1\pm k_2l=2\pi |n|.
\end{equation}
Here $k_1l\sin\theta_1$ is the difference of optical paths in half-space 1, $k_2l$ is the difference in half-space 2. The expression (\ref{synchronism}) is valid only for small diameters $2a\ll\lambda$.

Let us express the wavenumbers in terms of the wavelength
$k_{1,2}=2\pi\sqrt{\eps_{1,2}}/\lambda$. Eq. (\ref{synchronism}) can
be rewritten as
\begin{equation}\label{positive}
    \frac{l}{\lambda}=\frac{\pm|n|}{\sqrt{\eps_1}\sin\theta_1 \pm \sqrt{\eps_2}}.
\end{equation}
In vacuum for $n=1,2$ and $\theta_1=45^\circ$ it gives
$l/\lambda=0.5858, 1.1716, \dots$ Near the resonance a
rearrangement of the wave pattern occurs due to arising the next
order diffraction and transparency decreases sharply. The inset
demonstrates the splitting of the first dip. Its physical nature
is a difference between induced dipole moments along $x$ and $y$
directions. This difference is verified in the section
\ref{Dipole} within the coupled dipole approximation.

\begin{figure}
\psfrag{abs}{$l/\lambda$}
\psfrag{ord}{\hspace{-0.5cm}Transparency}
\includegraphics[width=\columnwidth]{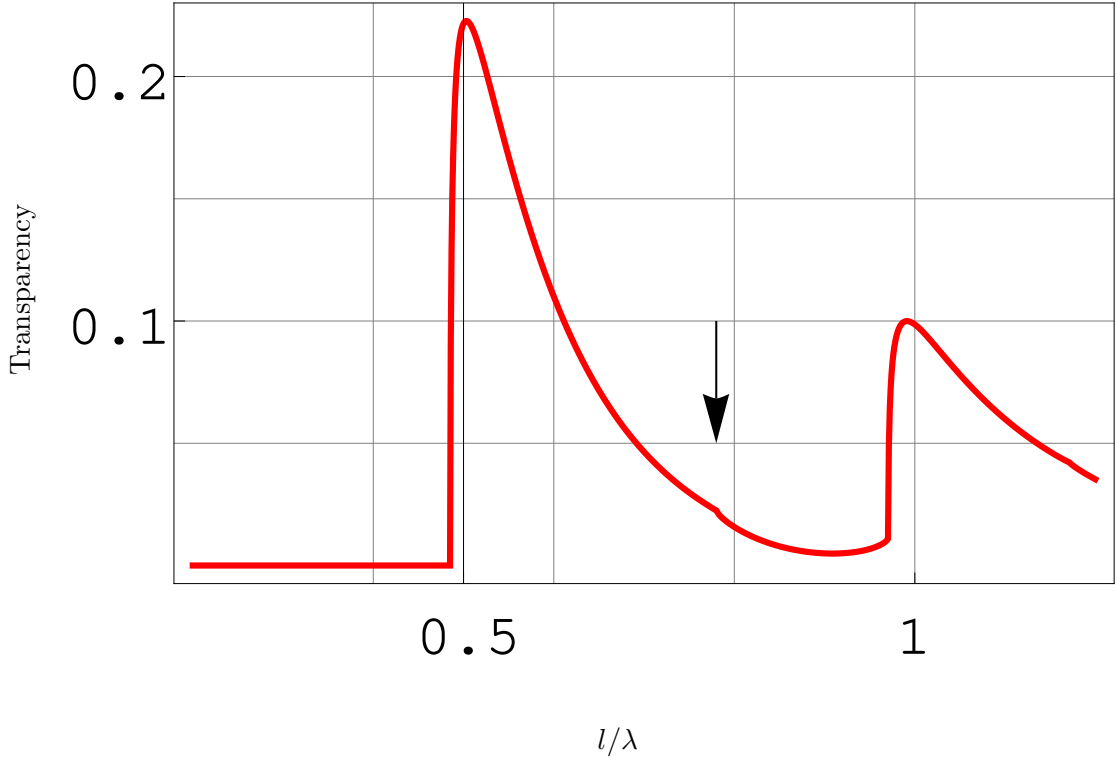}
\caption{The transparency as a function of period-to-wavelength ratio for evanescent wave ($\eps_1=2.25$). An arrow indicates the weak step.}\label{f:gold}
\end{figure}

The case of evanescent wave is shown in Fig. \ref{f:gold}. The
evanescent wave in the upper half space 2 arises from the
travelling wave falling onto the boundary by the angle of total
internal reflection (TIR). The refractive indices are
$\sqrt{\eps_1}=1.5, \sqrt{\eps_2}=1$ (describing a glass in the
air), the incident angle $\theta_1=45^\circ$, as above. The  limit
value of transparency at $l\gg\lambda$ is zero instead of unity.
The positions of resonances have shifted, since in  dielectric Eq.
(\ref{positive}) gives $l/\lambda=0.4853, 0.9706, \dots$

\begin{figure}
\psfrag{abs}{$l/\lambda$}
\psfrag{ord}{\hspace{-0.5cm}Transparency}
\includegraphics[width=\columnwidth]{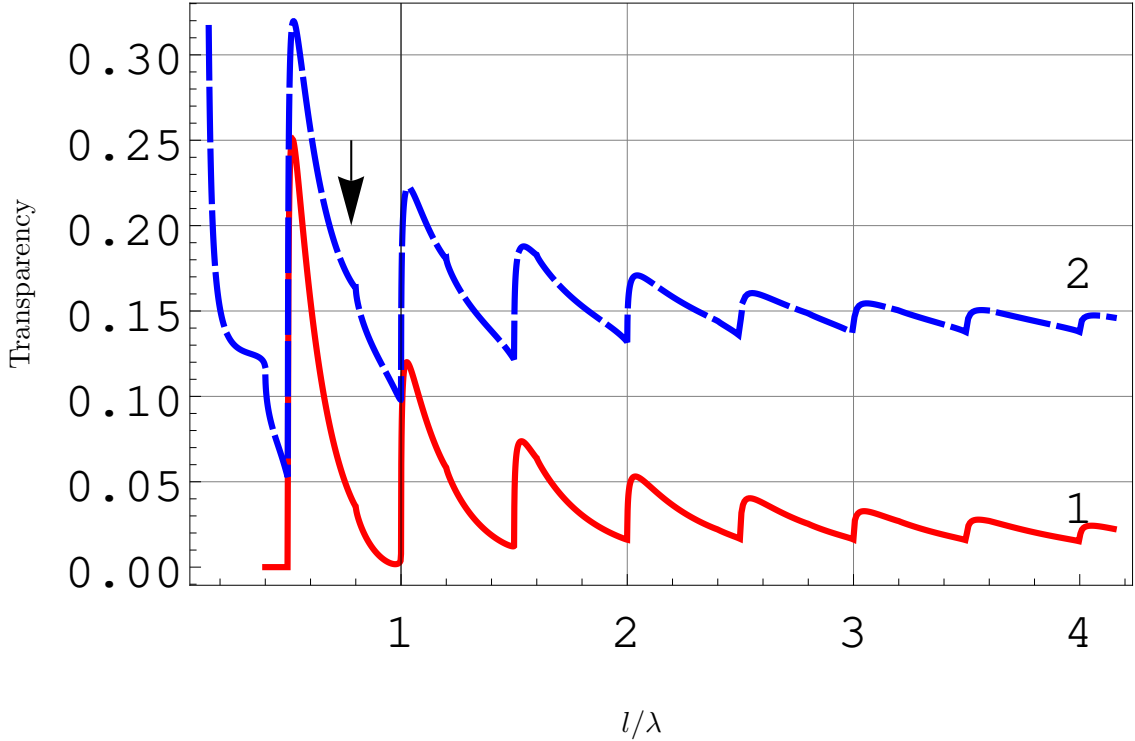}
\caption{The transparency as a function of period-to-wavelength ratio for a narrow gap $h=0.01\ \mu$m: the incidence angle is $\theta_1^{+}$ --- evanescent wave (curve 1) or $\theta_1^{-}$ --- traveling wave (curve 2). An arrow indicates the weak step.}\label{f:close-scattering}
\end{figure}

Comparing Fig. \ref{f:uniform} and \ref{f:gold} we see an
essential difference. There are four main distinctions of Fig.
\ref{f:gold}: the RWA broadening, the vanishing splitting, the
weak step (``jag'')  at $l/\lambda\approx0.78$, and the zero
asymptotic value at $l/\lambda\to\infty$. There are two
differences in conditions between Fig. \ref{f:uniform} and
\ref{f:gold}: the dielectric lower half-space with
$\sqrt{\eps_1}=1.5$ and the TIR incident angle. To clarify the
contributions of both differences let us repeat the calculation
for two very close angles, less and more than the TIR angle. We
fix $\eps_1=2.25, \eps_2=1$ and choose two values in incidence
angle $\theta_1$, both very close to the angle of total internal
reflection. For our case the TIR angle is equal to
$\theta_0=\arcsin\left(1/\sqrt{\eps_1}\right)=0.729728$. Our
values are $\theta_1^{-}\approx0.7297$ and
$\theta_1^{+}\approx0.7296$. First is less than TIR angle, the
second is greater. The difference is very small,
$|\theta_1^{\pm}-\theta_0|\approx10^{-4}$. At the same time the
angles corresponds the travelling or evanescent wave in medium 2
at almost equal conditions. The transparency curves are shown in
Fig.~\ref{f:close-scattering}. The upper curve, the transparency
for travelling wave, differs noticeably from the lower one.

When the angle $\theta_1$ is close to the TIR angle $\theta_0$,
only RWA of positive order exist at $l/\lambda=n/2, n=1,2,\dots$
The travelling wave propagates nearly along the surface at
$x$-direction and has mainly $y$-component of the electric field.
Then the $x$-component of the dipole moment is very small and the
splitting vanishes. The same situation takes place for angles
$\theta_1>\theta_0$, i.e. for evanescent wave, when $x$-component
of the dipole moment is exactly zero. Unlike, at a smaller angle
$\theta_1$ the $x$-component of dipole moment arises along with
the RWA splitting. Also, the broadening takes place for both the
curves.

\begin{figure}
\psfrag{abs}{$l/\lambda$}
\psfrag{ord}{\hspace{-0.5cm}Transparency}
\includegraphics[width=\columnwidth]{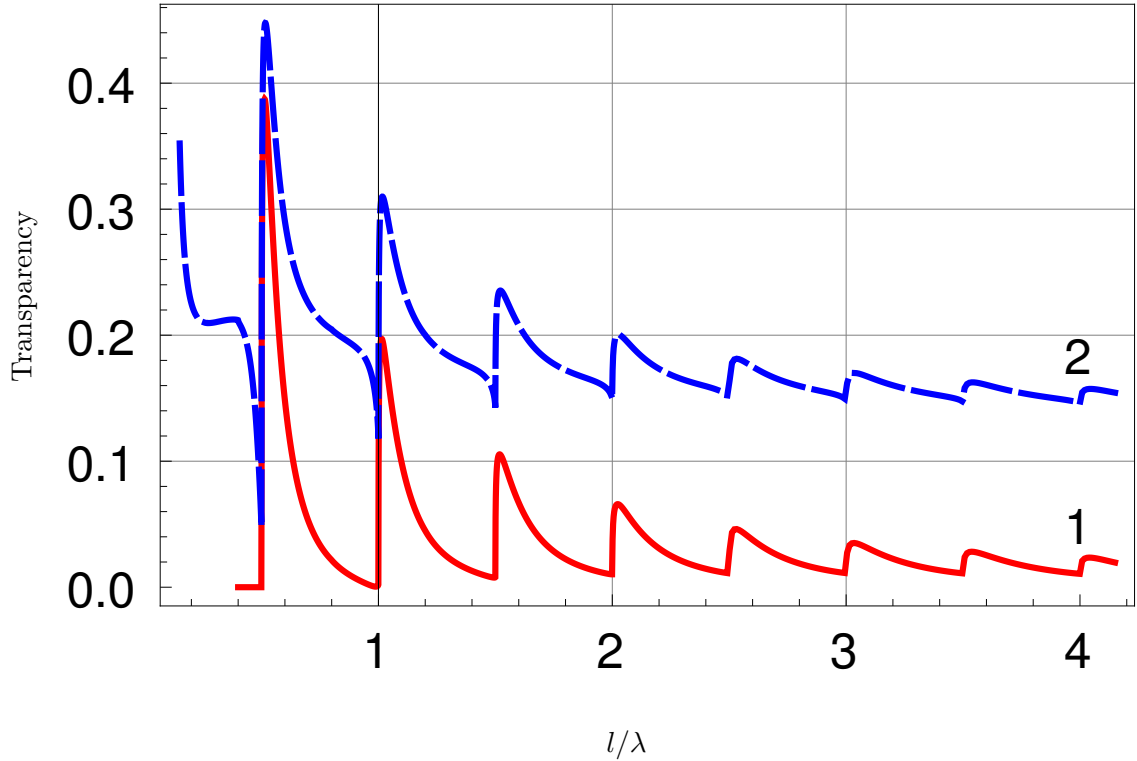}
\caption{The same as in Fig.~\ref{f:far-scattering} for the wider gap $h=0.3\ \mu$m.}\label{f:far-scattering}
\end{figure}

The picture considerably varies also with
changing the gap $h$. The computation in
Fig.~\ref{f:close-scattering} correspond to $h=10$~nm.
Fig.~\ref{f:far-scattering} shows the same curves for the
grating placed far from the boundary, at $h=300$~nm, when the
role of multiple reflection becomes relatively weak.
In comparison with  Fig.~\ref{f:close-scattering} the gap
increasing results in higher peak values of the transparency
coefficient. The step at $l/\lambda\approx0.78$
vanishes for wide gap, and then it
can be interpreted as an effect of the image source.

Thus the broadening and one component RWA are the effects of
dielectric. The jag is the result of image grating reflected
in the dielectric. The zero limiting value at $l/\lambda\to\infty$
is an effect of evanescent wave (zero background).

\subsection{Discrete Dipole Approximation}\label{Dipole}

In this section we consider the grid of nanowires within dipole
approximation. It means that we will describe the response of the
wire on the exciting electric field in terms of the wire's dipole
polarizability, $\alpha$. As before, we consider the case when the
electric field lies in $xy$-plane, the acquired dipole moments are
2-dimensional. Also, the dipole-dipole interaction between the
nanowires should be taken into account. Such an approach dealing
with {\em coupled} dipoles is widely used in nanophotonics,
particularly, in optics of nanoparticles
\cite{Purcell73,Markel91,Stockman97,Perminov03OiS}. The concept,
basically, consists in finding the values of the dipole moments,
acquired by the particles, when each is excited by both the
incident (given) field and the sum of the dipole fields from all
remaining particles. The dipoles obey the so-called Coupled
Dipoles Equation (CDE):
\begin{equation}
\label{CDE_general} \alpha_i^{-1} \mathbf{d}_i = \sum_{j\ne
i}{\hat{G}_{ij} \mathbf{d}_j} + \mathbf{E}_{0 i},
\end{equation}
where $\mathbf{E}_{0i} \equiv \mathbf{E}_0 (\mathbf{r}_i)$ --- the
electric field of the incident electromagnetic wave at the point of
$i$-th dipole's location; $\alpha_i$ is the polarizability of $i$-th
particle (the nanowire, in our case); summation is made over the
whole ensemble of the wires. The diadic $\hat{G}_{ij}$ is defined
through the electric field produced by $j$-th dipole at point
$\mathbf{r}_i$:
\begin{equation}
\label{G_def} E_\alpha(\mathbf{r}_i) = G_{i j, \alpha\beta }\, d_{j,
\beta},
\end{equation}
where Greek indices denote Cartesian components. Thus, $\hat{G}_{ij}
\mathbf{d}_j$ is the contribution from $j$-th dipole to the field,
acting on $i$-th one.

After the system of CDE has been solved, any quantity of interest
can be found using the values of $\mathbf{d}_i$. Here we calculate
the electric field near the nanowares array, to compare with the
results produced by BEM method.

\begin{figure}[htbp]
\psfrag{abs}{$l/\lambda$}
\psfrag{ord}{\hspace{-0.5cm}$|E/E_0|^2$}

\includegraphics[width=\columnwidth]{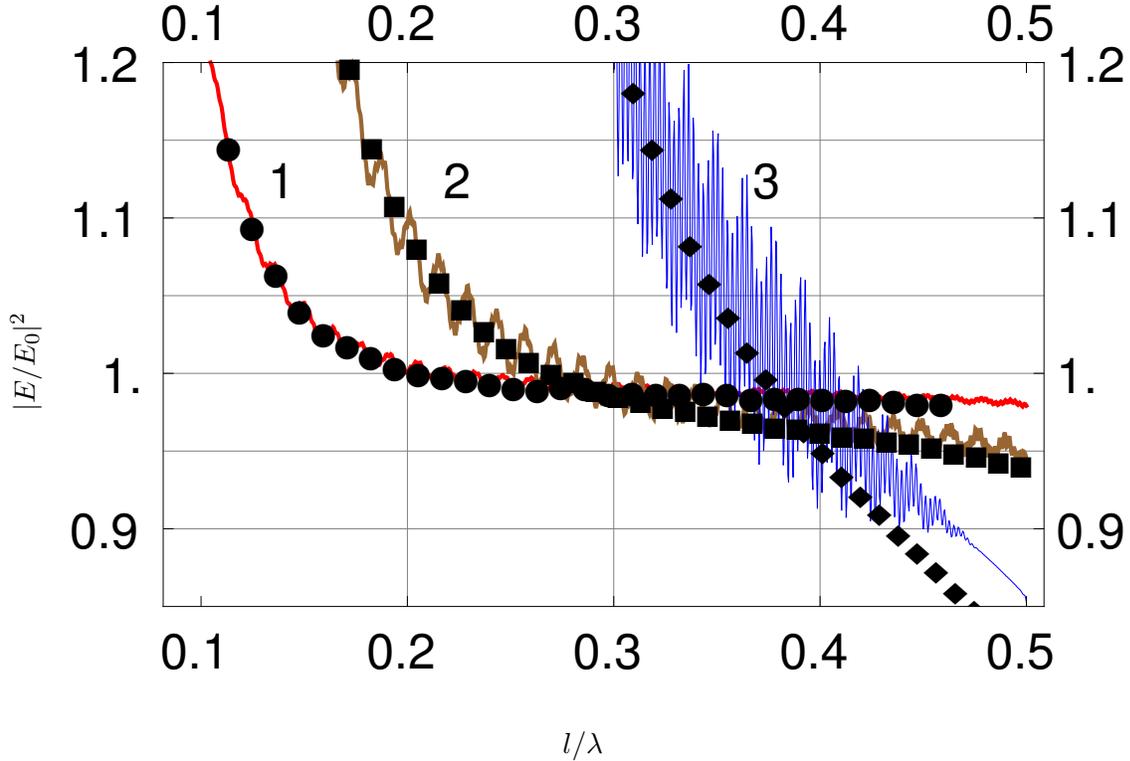}
\caption{Comparison of CDE solution for the chain of 500 2D
dipoles (solid lines) with that obtained by BEM for the infinite
array of nanocylinders (circles and squares). Plotted is the value
of squared electric field in the middle of the gap between the
cylinders (for CDE, between 250-th and 251-th ones). Here $a=30$
(curve 1), 50 (2), and 100~nm (3). The field is normalized by the
amplitude of the falling wave, $E_0$.} \label{CDE:f0}
\end{figure}

In Fig.\ref{CDE:f0} the squared modulus of the electric field,
calculated at the center of the chain of 500 gold nanowires, is
plotted for various spacing between them, $l$, and their radii, $a$.
The exciting field was the plane wave falling by the angle
$\theta_1= \pi/4$ from the left-bottom side (Fig.
\ref{fig:scheme}; the polarization is in $xy$ plane; the wavelength is
$\lambda = 1512$~nm.
As we see, the results of two compared methods perfectly match for
30-nm and 50-nm cylinders, while for the case of 100-nm radius
there is a noticeable difference. The discrepancy arise when
the diameter, $2a$, becomes comparable with
$\lambda / 2\pi$, destroying the dipole approximation.
Quite remarkable is that such a good agreement between
two methods remains up to the gap between the cylinders being as
small as their radius.

The computational difficulty of CDE solving rapidly grows with the
number of interacting dipoles. However, there is a special case
--- an infinite chain of dipoles, for which the calculations
become drastically more simple. It allows to consider the infinite
grating of nanowires analytically, at least for excitation by
plane travelling wave.

The main idea is that if the grating of nanowires and the aperture
of the incoming wave are both infinite, the wires become fully
equivalent. So, the difference in the corresponding dipole moments
may originate only from the different values of the incident field
at each dipole. In other words, the contribution to the total
(acting) field, which is due to dipole-dipole interaction,
$\sum_{j\ne i}{\hat{G}_{ij} \mathbf{d}_j}$, must have the same
amplitude at each $i$-th dipole, and must be always in phase with
the incoming field $\mathbf{E}_0(\mathbf{r}_i)$. Applying these
two conditions to (\ref{CDE_general}) we can get an analytical
solution of CDE. For this purpose, we use the explicit expression
for the electric field of an oscillating 2-dimensional dipole
moment $\mathbf{d} \propto \exp{(-i\omega t)}$
\begin{eqnarray}
\label{E_solut} \mathbf{E}(\mathbf{r}) = \frac{i \pi k }{r^3}
\Bigl[r^2 \mathbf{d}\left(kr \mathcal{H}^{(1)}_0(k r) -
\mathcal{H}^{(1)}_1(k r)\right)\nonumber\\ -
\mathbf{r}\left(\mathbf{d}\cdot\mathbf{r}\right)\left(kr
\mathcal{H}^{(1)}_0(k r) - 2 \mathcal{H}^{(1)}_1(k
r)\right)\Bigr],
\end{eqnarray}
which is derived using the approach, very similar to that
described in \cite{LL2} for 3-dimensional case. Also, we use the
asymptotic expression for both Hankel functions, that gives an
inaccuracy less than 15\% for $kr
> 0.5$. After summation we get:
\begin{eqnarray}
\label{Cde_Solution}d_x = E_0 \alpha \cos{\theta_1} \times\ \ \ \label{d_x} \\
\left\{1 - \frac{(1-i)\sqrt{\pi}\alpha }{8\sqrt{k
l^5}}[8 k l F_{3/2} + 3i F_{5/2}] \right\}^{-1},\nonumber \\
d_y = E_0 \alpha \sin{\theta_1} \times\ \ \ \label{d_y} \\
\left\{1 - \frac{(1+i)\sqrt{\pi}\alpha }{8\sqrt{k l^5}} [8 k^2 l^2
F_{1/2} + 7i k l F_{3/2} - 3 F_{5/2}] \right\}^{-1},\nonumber
\end{eqnarray}
where
\[
F_s \equiv \mbox{Li}_{s} (e^{i k l (1 - \sin{\theta_1})}) +
\mbox{Li}_{s} (e^{i k l (1 + \sin{\theta_1})}),
\]
$\mbox{Li}_{s}(z)$ is the polylogarithm \cite{Olver10}
\[
\mathrm{Li}_s(z)=\sum_{n=1}^\infty \frac{z^n}{n^s}.
\]
The dipole polarizability, $\alpha$, is the same for all the
nanowires; $k=\omega / c$.

\begin{figure}
\psfrag{abs}{$l/\lambda$}
\psfrag{ord}{\hspace{-1cm}$d_x\times10^{10},\ \textrm{SGSE}$}
\includegraphics[width=\columnwidth]{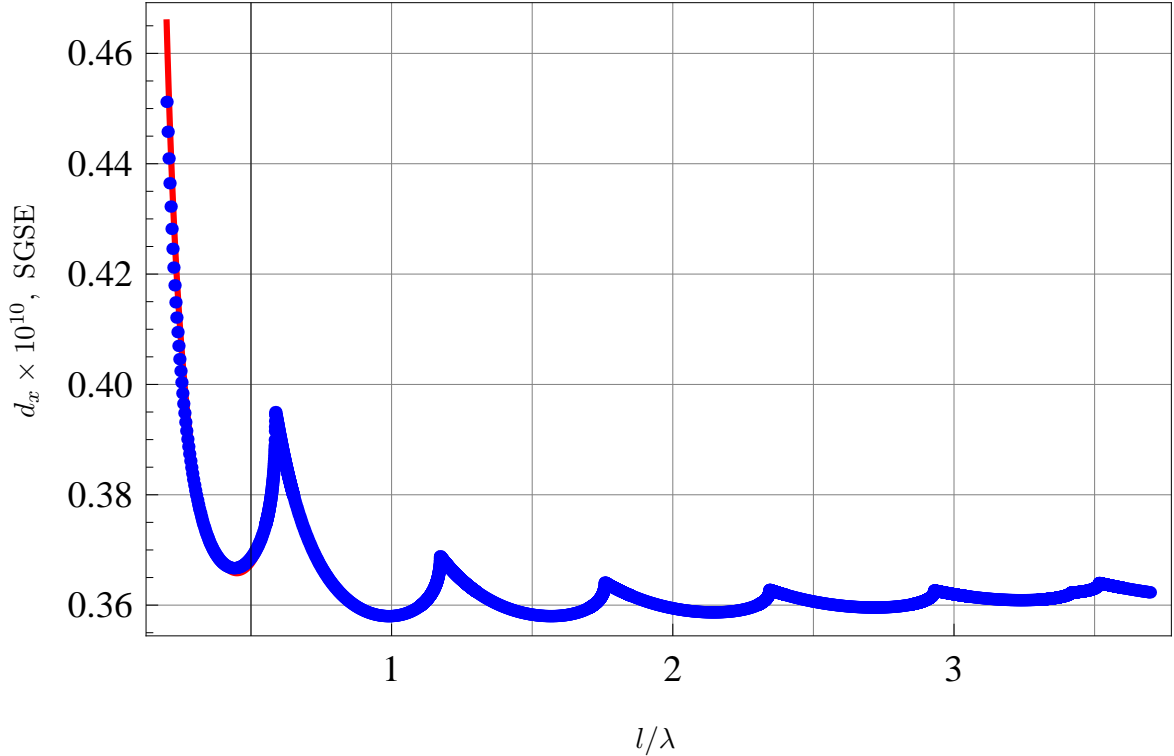}
\caption{The absolute value of x-components of the
dipole moment $d_x$ vs the relative spacing between the wires, $l /
\lambda$: the analytical formula for infinite number of wires (solid  line);
numerical calculation  for array of 500 elements (dots).} \label{CDE:f1}
\end{figure}

The resulting dipole moment $d_x$ is plotted in Fig.\ref{CDE:f1}.
The calculations were done for gold nanowires with the radius
100~nm as a dependence of the nanowires spacing, at constant
wavelength of incoming radiation $\lambda=1512$~nm. The
Rayleigh-Wood's anomalies are clearly seen in Fig.~\ref{CDE:f1} as
peaked extrema.

For comparison, we performed also a direct numerical solution of CDE
(\ref{CDE_general}) for 500 two-dimensional dipoles, formed by the
gold nanowires. The results for $d_x$ component (taken for the
``middle'' dipole, $i=250$) are in excellent quantitative agreement
with the analytical given above, as also shown in Fig.~\ref{CDE:f1}.
\begin{figure}
\psfrag{abs}{$l/\lambda$}
\psfrag{ord}{$\hspace{-1cm}d_x\times10^{10},\ \textrm{SGSE}$}
\includegraphics[width=\columnwidth]{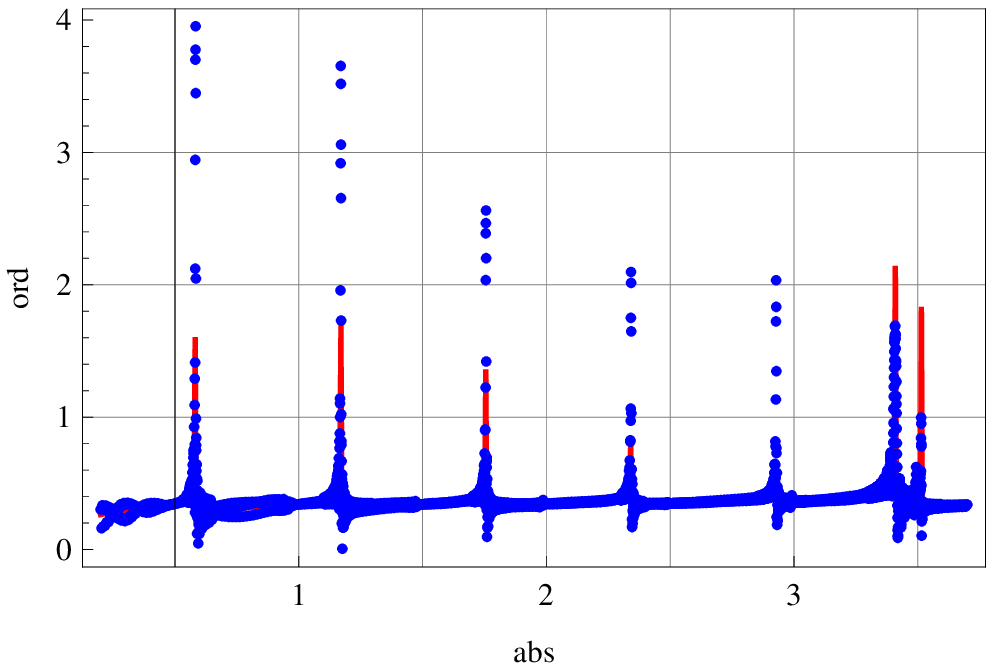}
\caption{The same as in Fig.\ref{CDE:f1} for $d_y$.
} \label{CDE:f2}
\end{figure}

The results of the same calculations for $d_y$ are shown in
Fig.~\ref{CDE:f2}. Comparing to Fig.~\ref{CDE:f1}, we see that
positions of peaks do not coincide with the case of $d_x$. It is
the shift that produces the splitting shown in
Fig.~\ref{f:uniform}. As for the amplitudes difference between
analytical and numerical solutions, we believe it to be a
consequence of high derivatives near the resonances, that drops
down the numerical accuracy.

\section{Conclusions}\label{Conclusion}

An evanescent wave  has to be considered together with its source.
The special compound Green function is proposed taking into
account the falling waves in the lower half space. Boundary
integral equations are extended for periodic infinite structure.
The effective Green function is
found allowing to treat an infinite periodic array of nanowires.
The numerical calculation for circular cylinders reveals Rayleigh
--- Wood's anomalies in the grating transparency. The dips turn
into peaks and splitting of resonances vanishes when total
internal reflection occurs. For the case of evanescent wave
nulling the asymptotic value of transparency is shown. The
calculations are compared with coupled dipole approximation using
the analytical solution, derived for an infinite number of coupled
dipoles, and the numerical (direct) calculation for the finite
grating with $N=500$ wires. In the latter case the oscillations
are found in the near-field, which are the result of a finite
number of cylinders.

\section*{Acknowledgement}

Authors are grateful to O. V. Belai and E. V. Podivilov for
helpful discussions. This work is supported by the Government
program of the leading research schools (NSh-2979.2012.2), program
\# 24 of the Russian Academy of Sciences Presidium, grant from the
Division of Physical Sciences, and Interdisciplinary grant \#68
from the Siberian Branch of RAS.

%

\end{document}